\documentclass[twocolumn]{aastex62}
\graphicspath{{./}{}}

\received{}
\revised{}
\accepted{}
\submitjournal{ApJL}

\shorttitle{Deep VLT photometry of YMCA-1.}
\shortauthors{M. Gatto et al.}

\begin{document}

\title{Deep VLT photometry of the faint stellar system in the Large Magellanic Cloud periphery YMCA-1.}

\correspondingauthor{M. Gatto}
\email{massimiliano.gatto@inaf.it}

\author[0000-0003-4636-6457]{Massimiliano Gatto}
\affiliation{INAF--Osservatorio Astronomico di Capodimonte, Via Moiariello
 16, 80131, Naples, Italy \\}
\affiliation{Dept. of Physics, University of Naples Federico II, C.U. Monte Sant'Angelo, Via Cinthia, 80126, Naples, Italy\\}

\author{V. Ripepi}
\affiliation{INAF-Osservatorio Astronomico di Capodimonte, Via Moiariello
 16, 80131, Naples, Italy \\}
 
\author[0000-0001-8200-810X]{M. Bellazzini}
\affiliation{INAF-Osservatorio di Astrofisica e Scienza dello Spazio, Via Gobetti 93/3, I-40129 Bologna, Italy}

\author{M. Dall'ora}
\affiliation{INAF-Osservatorio Astronomico di Capodimonte, Via Moiariello 16, 80131, Naples, Italy}

\author{M. Tosi}
\affiliation{INAF-Osservatorio di Astrofisica e Scienza dello Spazio, Via Gobetti 93/3, I-40129 Bologna, Italy}

\author{C. Tortora}
\affiliation{INAF-Osservatorio Astronomico di Capodimonte, Via Moiariello 16, 80131, Naples, Italy}

\author{M. Cignoni}
\affiliation{INAF-Osservatorio Astronomico di Capodimonte, Via Moiariello 16, 80131, Naples, Italy}
\affiliation{Physics Departement, University of Pisa, Largo Bruno Pontecorvo, 3, I-56127 Pisa, Italy}
\affiliation{INFN, Largo B. Pontecorvo 3, 56127, Pisa, Italy}

\author{M.-R. L. Cioni}
\affiliation{Leibniz-Institut f\"ur Astrophysik Potsdam, An der Sternwarte 16, D-14482 Potsdam, Germany}

\author{F. Cusano}
\affiliation{INAF-Osservatorio di Astrofisica e Scienza dello Spazio, Via Gobetti 93/3, I-40129 Bologna, Italy}

\author{G. Longo}
\affiliation{Dept. of Physics, University of Naples Federico II, C.U. Monte Sant'Angelo, Via Cinthia, 80126, Naples, Italy}

\author{M. Marconi}
\affiliation{INAF-Osservatorio Astronomico di Capodimonte, Via Moiariello 16, 80131, Naples, Italy}

\author{I. Musella}
\affiliation{INAF-Osservatorio Astronomico di Capodimonte, Via Moiariello 16, 80131, Naples, Italy}

\author{P. Schipani}
\affiliation{INAF-Osservatorio Astronomico di Capodimonte, Via Moiariello 16, 80131, Naples, Italy}

\author[0000-0002-6427-7039]{M. Spavone}
\affiliation{INAF-Osservatorio Astronomico di Capodimonte, Via Moiariello 16, 80131, Naples, Italy}





\begin{abstract}
We present FORS2@VLT follow-up photometry of YMCA-1, a recently discovered stellar system located 13\degr~from the Large Magellanic Cloud (LMC) centre.
The deep colour-magnitude diagram (CMD) reveals a well-defined main-sequence (MS) and a handful of stars in the post-MS evolutionary phases.
We analyse the YMCA-1 CMD by means of the automated isochrone matching package {\tt ASteCA} and model its radial density profile with a Plummer function. We find that YMCA-1 is an old ($11.7^{+1.7}_{-1.3}$~Gyr), metal-intermediate ([Fe/H] $\simeq -1.12^{+0.21}_{-0.13}$~dex), compact (r$_{\rm h} = 3.5 \pm 0.5$ pc), low-mass (M $= 10^{2.45 \pm 0.02} M_{\odot}$) and low-luminosity (M$_V = -0.47 \pm 0.57$~mag) stellar system.
The estimated distance modulus ($\mu_0 = 18.72^{+0.15}_{-0.17}$~mag), corresponding to about 55~kpc, suggests that YMCA-1 is associated to the LMC, but we cannot discard the scenario in which it is 
a Milky Way satellite.
The structural parameters of YMCA-1 are remarkably different compared with those of the 15 known old LMC globular clusters. In particular, it resides in a transition region of the M$_V$-r$_h$ plane, in between the ultra-faint dwarf galaxies and the classical old clusters, and close to SMASH-1, another faint stellar system recently discovered in the LMC surroundings.
\end{abstract}



\section{Introduction}

One of the most lively fields of modern astrophysics is the search for faint stellar systems inhabiting the Milky Way (MW) halo or the periphery of the MW satellites.
Amongst the tens of MW satellites, the Large Magellanic Cloud (LMC) has particular relevance, as it is the largest, and is known to have entered the MW halo with its own system of dwarf galaxy satellites \citep[][]{Kallivayalil-2018}.
The LMC is also known to possess at least 15 globular clusters (GCs), as old as its oldest stars (i.e. 12-13 Gyrs), which can be used to probe the earliest phases of its evolution.
For example, based on accurate spectroscopic analysis, \citet{Mucciarelli-2021} discovered that the old cluster NGC~2005 has been captured by the LMC from a smaller satellite galaxy now completely dissolved.\par
In recent years, thanks to the advent of deep panoramic surveys probing large portions of the sky, the number of faint stellar systems discovered in the vicinity of the Magellanic Clouds (MCs) has dramatically increased \citep{Koposov-2015,Bechtol-2015,Drlica-Wagner-2015,Kim&Jerjen2015,Torrealba-2018,Martin-2016a,Bellazzini-2019}, bringing new puzzle pieces to the reconstruction of the MCs evolutionary history.
For example, \citet{Martin-2016a} reported the discovery of SMASH-1, a faint stellar system  at 11.3\degr~in projection from the LMC centre, 
whose properties place it in between the classical GCs and the ultra-faint dwarf galaxies (UFDs).
The compactness of SMASH-1 led the authors to suggest that it likely is an old star cluster (SC) fundamentally different from UFDs which are heavily dark-matter dominated \citep[see e.g.,][and references therein]{Simon2019}.
Nonetheless, the properties of SMASH-1 are very different from those shown by the historically known LMC old GCs. Indeed, unlike these objects, it is faint ($L_{\rm V} = 10^{2.3} L_{\odot}$), compact ($r_{\rm h} = 9.1$~pc), and highly elliptical in shape.
Detecting these old stellar systems and unveiling their origin is of primary importance to understanding how the MCs and galaxies in general form and evolve.
Here we discuss another stellar system, similar to SMASH-1, that we identified for the first time through the survey ''Yes, Magellanic Clouds Again'' (YMCA, PI: V. Ripepi) and dubbed YMCA-1 \citep[][]{Gatto-2021a}. YMCA is carried out with the VLT Survey Telescope \citep[VST;][]{Capaccioli&Schipani2011}. One of the main objectives of the survey is to discover faint stellar systems in the MC peripheries. To this aim, we performed an extensive search of unknown SCs in the periphery of the LMC, by means of an automated algorithm which looks for over-densities in the sky  \citep[see][]{Gatto-2020}. YMCA-1 is located at about 13\degr~to the East of the LMC centre \citep[][]{Gatto-2021a}.
The analysis of the color-magnitude diagram (CMD) of YMCA-1 based on VST data, carried out by means of visual isochrone fitting, suggested that it is an old (t$>$12 Gyr) and metal-poor ([Fe/H]$\sim$-2.0 dex) stellar system, while its estimated distance (D$\sim$100 kpc) placed YMCA-1 in the outermost regions of the MW halo \citep[see][]{Gatto-2021a}.
However, the available VST data, which revealed only a few stars in the red-giant branch (RGB) and in the top main sequence (MS), were not sufficiently deep to unambiguously establish its real physical nature, mainly because we could not obtain a robust distance for the target.
Indeed, given the lack of evolved distance indicators such as Horizontal Branch (HB) or Red Clump (RC) stars, the distance can only be constrained by a clear identification of the MS of the system. This can only be achieved with deep follow-up photometry.
Hence, to unveil the real nature of this very interesting stellar system, we carried out deep follow-up photometry with the ESO (European Southern Observatory) very large telescope (VLT). In this letter, we report and discuss the results obtained for YMCA-1 based on this new deep data.

\section{Observations and data reduction}

Deep photometric data for YMCA-1 were obtained with the FORS2 imager of the VLT.
The observations were carried out during the nights of November 2 and 29, 2021, for the $g_{\rm HIGH}$ and $I_{\rm BESSEL}$ filters, respectively. The observations were divided into 5 sub-exposures 480s each in the $g_{\rm HIGH}$ filter and 13 sub-exposures of 240s each in the $I_{\rm BESSEL}$-band, to reach faint magnitudes without saturating the bright members of YMCA-1. The typical seeing was of 0.51$''$ and 0.72$''$ in $g$ and $I$ respectively.  
For the setup of FORS2, we chose a pixel scale of 0.25 arcsec/pixel with a field of view $6.8'~\times~6.8'$. FORS2 is equipped with a mosaic of two 2k$\times$4k MIT CCDs\footnote{see the manual at http://www.eso.org/sci/facilities/paranal/ instruments/fors/doc/VLT-MAN-ESO-13100-1543\_P01.pdf}. As the dimension of YMCA-1 is much smaller than the FoV of each of the two CCDs of FORS2, we decided to place the target only on the top CCD which has a larger FoV than the bottom one. We adopted a dithering procedure between the different sub-exposures to eliminate cosmic rays and bad pixels. 
The images were pre-reduced (de-biasing and flat-fielding) using the standard procedures with the {\tt IRAF} package \citep{IRAF-1-1986,IRAF-2-1993}. 
To obtain the photometry, we adopted the DAOPHOT/ALLFRAME packages \citep{Stetson1987,Stetson1994} which are best suited to reach faint magnitudes in a relatively crowded field such as YMCA-1. In brief, the different steps of the procedure were the following:
\begin{itemize}
\item A quadratically varying PSF was modelled, by letting the code free to adopt the function which minimized the $\chi^2$ of the fit. The most used function was the Moffat25, while in some cases the alogorithm choosed the Penny1 or the Penny2 functions. A WCS plate solution was computed for each individual image by querying the astroquery.astrometry\_net python module. Then, stars' XY position were converted to WCS coordinates by using the WCSCTRAN command, available under IRAF.
\item A stack of all the sub-exposures was created with MONTAGE2 \citep{Stetson1987,Stetson1994} to obtain a master list of sources on the image as deep as possible. 
\item ALLFRAME was run on all the sub-exposures using the derived master list as input for the stars' position
\item DAOMATCH/DAOMASTER \citep{Stetson1987,Stetson1994} were used to match the 5 and 13 different photometric catalogues obtained for each exposure of the $g_{\rm HIGH}$ and $I_{\rm BESSEL}$ filters. Finally, the catalogue in the two bands was put together. 
\end{itemize}
The absolute photometric calibration was obtained by means of the stars provided by the stars in common with the VST catalogue of the tile in which YMCA-1 resides, namely the tile YMCA 9\_47.
In particular, we cross-matched the PSF photometric catalog with the VST data by adopting a search radius of 0.5$''$. Then we corrected for the colour dependence of the zero points in $g$ and $i$ filters\footnote{We obtained the following calibration equations: $g = g_{\rm HIGH} + 9.656 \pm 0.005 -(0.033 \pm 0.004) (g_{\rm HIGH} - I_{\rm BESSEL})$ 
and $i = I_{\rm BESSEL} + 8.675 \pm 0.004 + (0.0946 \pm 0.003) (g_{\rm HIGH} - I_{\rm BESSEL})$.}.
Before exploiting the YMCA-1 photometric catalogue obtained as described above, we applied a cleaning procedure to remove undesired extended sources and the remaining few spurious detections. To this aim, we used the {\sc SHARPNESS} parameter of the {\it DAOPHOT} package, retaining only sources having $-0.15 < SHARPNESS < 0.15$.

\section{Analysis}

\begin{figure*}
    \includegraphics[width=0.32\hsize]{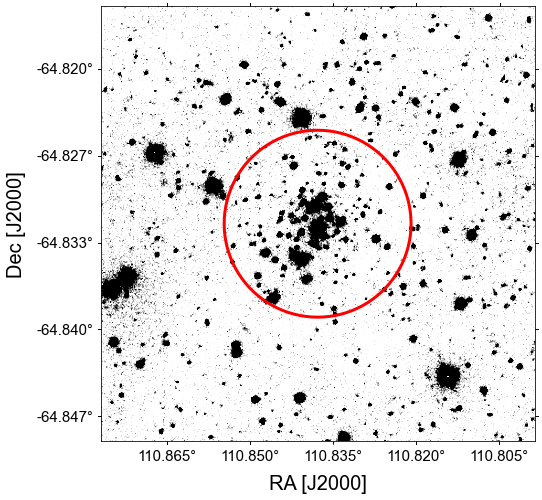}
    \includegraphics[width=0.37\hsize]{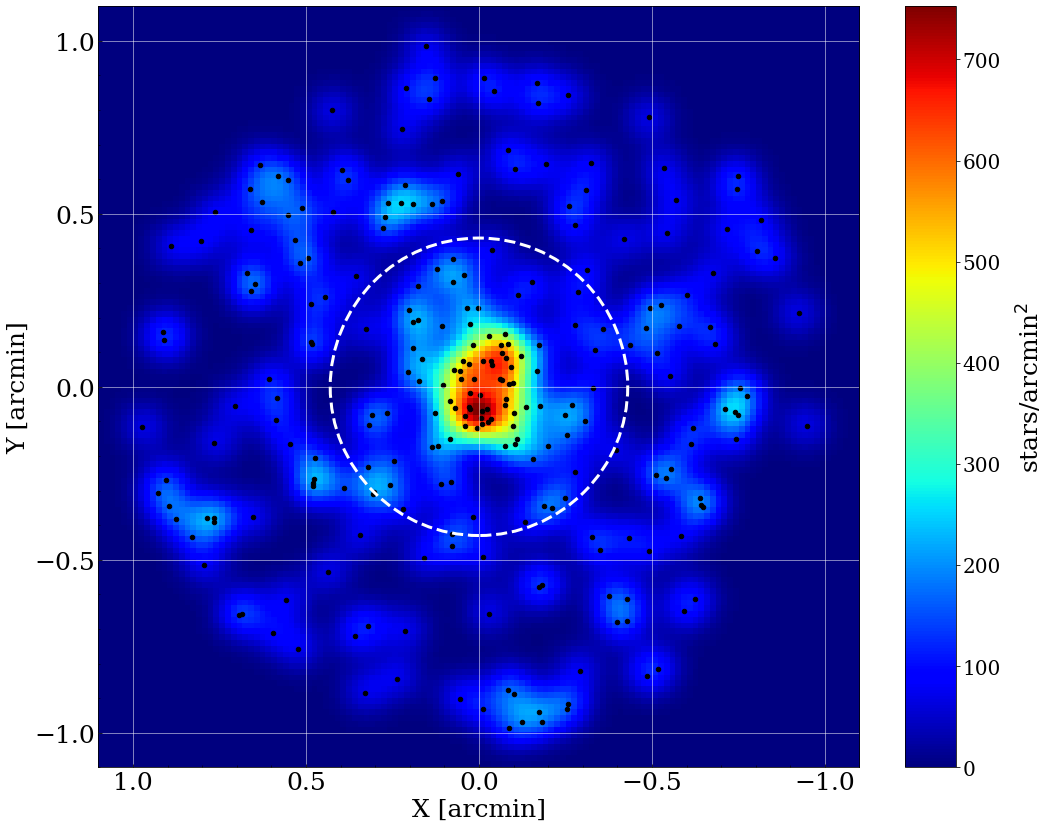}
    \includegraphics[width=0.32\hsize]{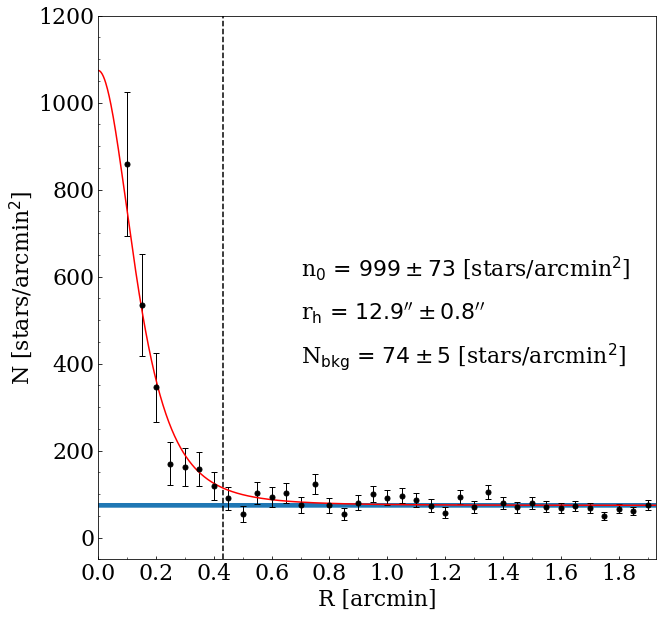}
    \caption{\emph{Left:} Sky image of a region of radius r = 1\arcmin~around the YMCA-1 centre. The red circle indicates the area defined by $2r_{\rm h} = 0.43\arcmin$.
    \emph{Centre:} Density map of stars' relative positions with respect to the YMCA-1 centre in a circular region of 1\arcmin~in radius.  We used a Gaussian function with $\sigma = 0.05\arcmin$ to smooth the map. Black points indicate the position of the stars, while the white dashed circle marks a radius r = 0.43\arcmin, namely twice the estimated half-light radius.
    \emph{Right:} Radial density profile of YMCA-1. Each point represents the density of stars in shells having a radius of 0.05\arcmin. Errors are Poissonian. The red solid line is a best fit of a Plummer model as indicated in the text, whose parameters are indicated at the centre of the figure. The horizontal blue strip region marks the N$_{\rm bkg} \pm 1\sigma$ estimated values.
    The vertical dashed line is at r = 0.43\arcmin, namely 2$r_{\rm h}$.}
    \label{fig:ymca1-position}
\end{figure*}


\begin{figure*}
    \centering
    \includegraphics[width=\hsize]{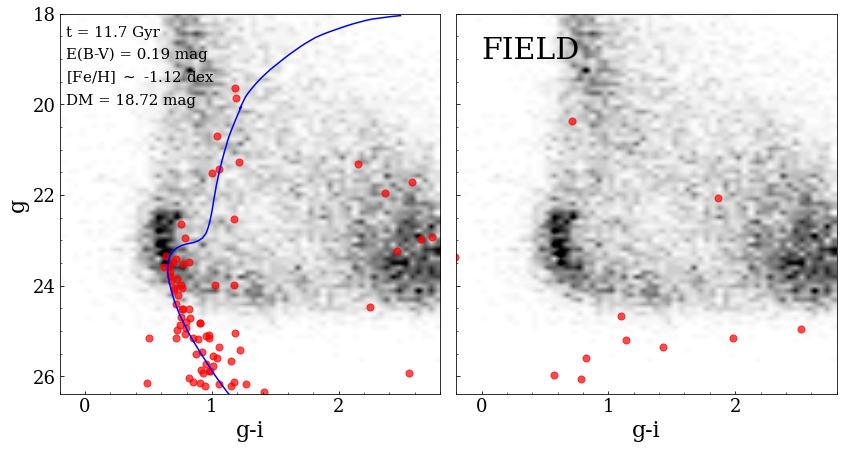}
    \caption{\emph{Left: } Stars whose photometry was obtained with the VLT (red points) within r$_{\rm h} = 0.43\arcmin$ from YMCA-1 centre. In the background, as grey points, the stars whose photometry was obtained with the VST and within 30\arcmin~from YMCA-1 centre. The blue solid line represents the best isochrone found with the ASteCA python package matching YMCA-1 stars, whose parameters are reported in the left top corner. \emph{Right:} Same as the left panel, but the red points are stars of a representative local field, which is a shell having an inner radius of 1\arcmin~ and outer radius set in order to have the same area adopted in the left panel.}
    \label{fig:ymca1_cmd_VLT}
\end{figure*}

The first step of the analysis consisted in estimating the centre of YMCA-1 by means of a technique based on Kernel Density Estimation (KDE) algorithm \citep[see sect. 3.1 in][for full details]{Gatto2021b}. As a result, the coordinates (J2000) of YMCA-1 centre are (RA, Dec) = (110.8378\degr, -64.8319\degr).
The left panel of Fig.~\ref{fig:ymca1-position} shows a sky map of the region (radius = 1\arcmin) around the YMCA-1 centre, while the central panel displays a density map of the same region, smoothed by means of a Gaussian function with bandwidth = 0.05\arcmin. 
There is a remarkable over-density of stars with respect to the YMCA-1's surroundings, clearly asserting the presence of a stellar system.
These maps also suggest that YMCA-1 might be slight elongated in the North-West South-East direction. To explore this possibility, we adopted the same method used by \citet{Martin-2016b} to estimate (among the other parameters) the ellipticity and position angle of SMASH-1 \citep[][]{Martin-2016a}. However, the resulting probability density function of the ellipticity and position angle were unconstrained and therefore in the following we employ a conservative approach, neglecting any possible elongation in the analysis of YMCA-1.
The right panel of Fig.~\ref{fig:ymca1-position} displays the radial density profile (RDP, number of stars per squared arcmin) of YMCA-1 built by using concentric shells of 0.05\arcmin~size. 
We modeled the RDP with a \citet{Plummer1911}'s profile, whose analytic form is the following:
\begin{equation}
    n(r) =  \frac{n_0 \cdot r_{\rm h}^4}{(r_{\rm h}^2+r^2)^2} + N_{\rm bkg}
\end{equation}
\noindent
where n$_0$ is the central surface density, r$_{\rm h}$ is the half-light radius, and N$_{\rm bkg}$ is the estimated level of the background density.
The free parameters of the model are n$_0$, r$_{\rm h}$ and N$_{\rm bkg}$. 
To fit the model to the data we adopted the {\tt curve\_fit} routine of the {\tt scipy} python library, which allows us to estimate the best parameters of the Plummer's model through a non-linear least-squares method.
The outcomes of the fit are labeled in Fig.~\ref{fig:ymca1-position} (right panel). In particular, the half-light radius $r_{\rm h}$ is an important parameter to select in an objective way stars likely members of YMCA-1.\par
In Fig.~\ref{fig:ymca1_cmd_VLT} (left panel) we display the CMD of YMCA-1 within a radius R=$0.43\arcmin$ around its centre, corresponding to twice the half-light radius $r_{\rm h}$. 
The right panel of Fig.~\ref{fig:ymca1_cmd_VLT} displays the CMD of a representative local field, taken at $1\arcmin$ from YMCA-1 centre and having an area as large as the area adopted in the left panel. 
In the background of both panels of the same figure, we show the stars observed with the VST in a region of $30\arcmin$ around YMCA-1. 
The CMD of YMCA-1 shows a well-defined MS, which extends below $g$=26 mag, that is at least 2.5 mag below the Turn-Off (TO), while the morphology for $g<$24 mag is similar to that depicted by the VST data. The comparison with the field provides further confirmation about the physical reality of YMCA-1. To exploit our deep CMD we adopted the Automated Stellar Cluster Analysis package \citep[\tt ASteCA,][]{Perren-2015} which allows us to perform an automated search of the best isochrone model which matches the data.
In particular, {\tt ASteCA} compares the position of the stars in the CMD with those of synthetic generated single stellar populations (SSP), adopting a genetic algorithm to find the best solution \citep[see][for full details]{Perren-2015}. We ran {\tt ASteCA} on a set of PARSEC isochrone models \citep[][]{Bressan-2012} to estimate age, reddening, metallicity, distance modulus and mass values of YMCA-1 and their uncertainties. To speed up the operations with the {\tt ASteCA} package, we feed it with realistic priors, namely: t $\geq$ 10 Gyrs, E(B-V) $\leq$ 0.3 mag, 10$^{-3} \leq$ Z $\leq 10^{-1}$ and 18.0 $\leq$ (m - M) $\leq$ 19.50 mag. The results of the application of {\tt ASteCA}, are listed in Table~\ref{tab:ymca-1_properties}. The isochrone with the best fitting parameters is overlaid on the data in the left panel of Fig.~\ref{fig:ymca1_cmd_VLT}.
The {\tt ASteCA} fit provides a distance modulus $\mu_0 = 18.72^{+0.15}_{-0.17}$~mag which corresponds to about 55~kpc, which is a significantly smaller value compared to the $\sim$100 kpc estimated with the shallower VST photometry \citep{Gatto-2021a}.    
Similarly to the analysis with VST data, we find that the age of YMCA-1 is $t \sim 11.7^{+1.7}_{-1.3}$~Gyr, but with a higher metallicity ([Fe/H] $\simeq -1.12^{+0.21}_{-0.13}$~dex)\footnote{We adopted the PARSEC Z$_{\odot}$ = 0.0152 value.}. It is confirmed that YMCA-1 is a 
compact (r$_{\rm h} = 3.5 \pm 0.5$ pc) stellar system.\par
Figure~\ref{fig:ymca1_cmd_VLT} also shows that the average main-sequence turn-off (MSTO) of the LMC field stars, shown as grey points in the CMDs, seems to be brighter compared to that of YMCA-1. As the LMC stellar population at the outer rim of the LMC disc should also be old and metal-poor \citep[][]{Mazzi-2021}, we speculate that the magnitude difference between the LMC and YMCA-1 MSTOs raises from a different distance modulus.
The currently adopted distance for the LMC centre is $\sim 49.6$ kpc \citep[e.g.][]{Pietrzynski-2019}, which corresponds to a distance modulus of DM $\sim 18.49$ mag, but the LMC disc is inclined in such a way that the North-East side (i.e. where YMCA-1 resides) is closer to us \citep[e.g.,][]{Choi-2018a}\footnote{There is any definitive evaluation of how closer to us the LMC is at YMCA-1 distance.}.
Therefore, YMCA-1 should be placed well behind the LMC main disc.

\section{Discussion}

The estimated distance of YMCA-1 suggests that it is likely associated to the LMC. Indeed, its three-dimensional distance from the LMC is $\sim 13$ kpc, well within the LMC tidal radius \citep[i.e. $\sim$16 kpc measured by][]{vanderMarel&Kallivayalil2014}. However, the possibility that YMCA-1 is incidentally projected beyond the LMC but not physically associated to this galaxy cannot be ruled out yet, as we still lack radial velocity measurements of its member stars.\par%
To further investigate the YMCA-1 properties, it is useful to compare them with those of SMASH-1, which appears to have close similarities with YMCA-1. 
To this aim, we first estimate the total luminosity and the stellar mass of YMCA-1 with a technique similar to that described in \citet{Gatto-2021a}. In brief, we adopted a synthetic SSP with $t \simeq 11.7$~Gyr and [Fe/H] $\simeq -1.12$~dex (corresponding to the best isochrone found with {\tt ASteCA}) constructed by means of the PARSEC isochrones\footnote{http://stev.oapd.inaf.it/cgi-bin/cmd}. Then, we measure the total luminosity and total mass of the synthetic SSP with a comparable number of MS stars as observed in YMCA-1.
In particular, we consider only YMCA-1 MS stars in the magnitude interval $23.5 \leq g \leq 25$ and with a maximum colour distance of 0.2 mag from the best isochrone (i.e. $24 \pm 5$ YMCA-1 stars by adopting the estimated r$_{\rm h} = 0.43\arcmin$). The bright limit was set to select only MS stars, avoiding the use of the much less populated sub-giant branch (SGB) and RGB phases.
The faint magnitude limit, instead, was chosen to take into account that at a fainter level the completeness problems could become significant.
After 500 random extractions we estimated a total luminosity for YMCA-1 equal to $L_g = 10^{2.1 \pm 0.3} L_{\odot}$ and $L_i = 10^{2.1 \pm 0.4} L_{\odot}$ and a total mass of M $= 10^{2.45 \pm 0.02} M_{\odot}$.\par
\begin{table}
    \caption{Properties of YMCA-1.}
    \label{tab:ymca-1_properties}
    \large
    \begin{tabular}{l|l}
    \hline \hline \smallskip
    Property & Value \\
    \hline \smallskip
    RA (J2000) & 110.8378\degr\\
    Dec (J2000) & -64.8319\degr\\
    Age & $11.7^{+1.7}_{-1.3}$~Gyr\\
    $\mu_0$ & $18.72^{+0.15}_{-0.17}$~mag\\
    $[$Fe/H$]$ & $-1.12^{+0.21}_{-0.13}$~dex\\
    E(B-V) & $0.19^{+0.04}_{-0.02}$~mag\\
    Mass & $10^{2.45 \pm 0.02} M_{\odot}$\\
    $r_h$ & $3.5 \pm 0.5$~pc\\
    $L_g$ & $10^{2.1 \pm 0.3} L_{\odot}$\\
    $L_i$ & $10^{2.1 \pm 0.4} L_{\odot}$\\
    $M_g$ & $-0.18 \pm 0.50$ mag\\
    $M_i$ & $-0.83 \pm 0.55$ mag\\
    $M_V$ & $-0.47 \pm 0.57$ mag\\
    \hline
    \end{tabular}
\end{table}

\begin{figure*}
    \centering
    \includegraphics[width=\textwidth]{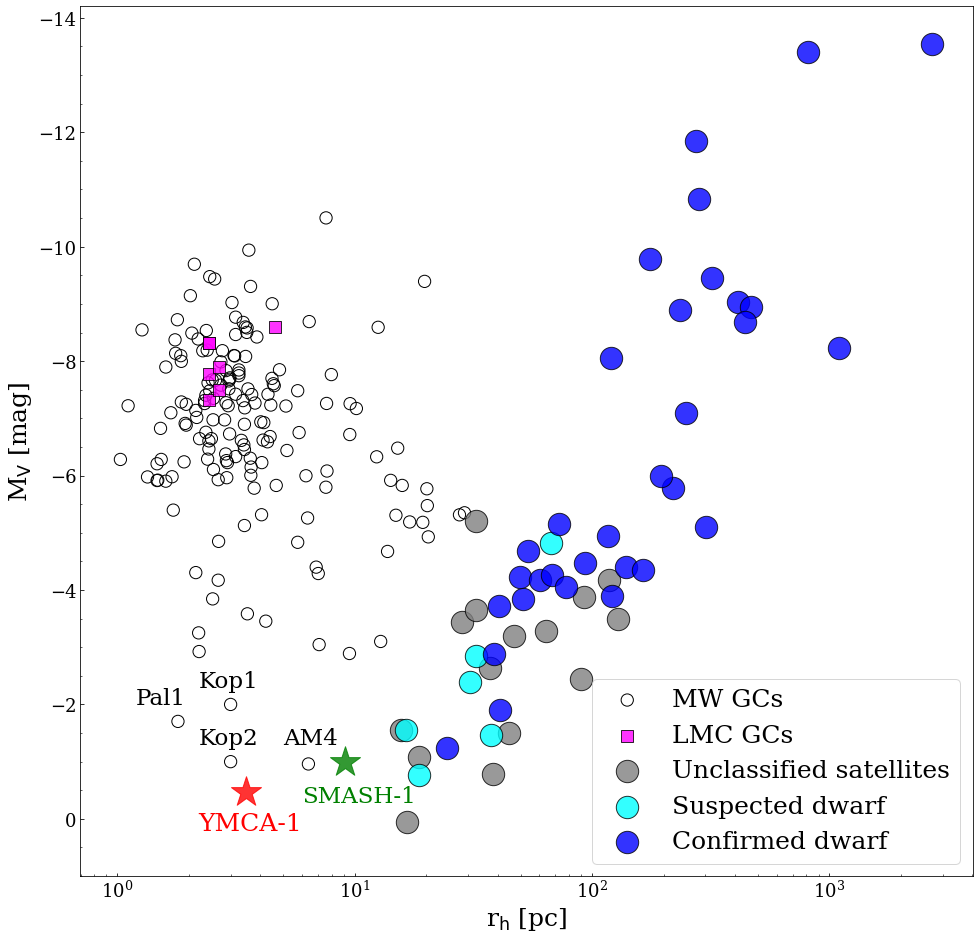}
    \caption{M$_{\rm V}$ versus r$_{\rm h}$ in which we depict the position in this plane of YMCA-1 and SMASH-1 (coloured stars), of some old LMC GCs (magenta squares; r$_{\rm h}$ taken from \citealt{Piatti&Mackey2018} and M$_{\rm V}$ taken from \citealt{Mackey&Gilmore2003a}) and MW GCs (empty circles;  taken from the \citet{Baumgardt&Hilker2018} catalog and from \citealt{Koposov-2007}). We retrieved reddening values from \citealt{Harris1996}, 2010 version, with some exceptions as listed in Tab.1 in \citealt{Gatto-2021a}). Finally, we also indicate the position of some confirmed and probable dwarfs (coloured circles) reported in the Supplementary Table 1 in \citealt{Simon2019}.}
    \label{fig:simon_image}
\end{figure*}

\begin{figure}
    \centering
    \includegraphics[width=\hsize]{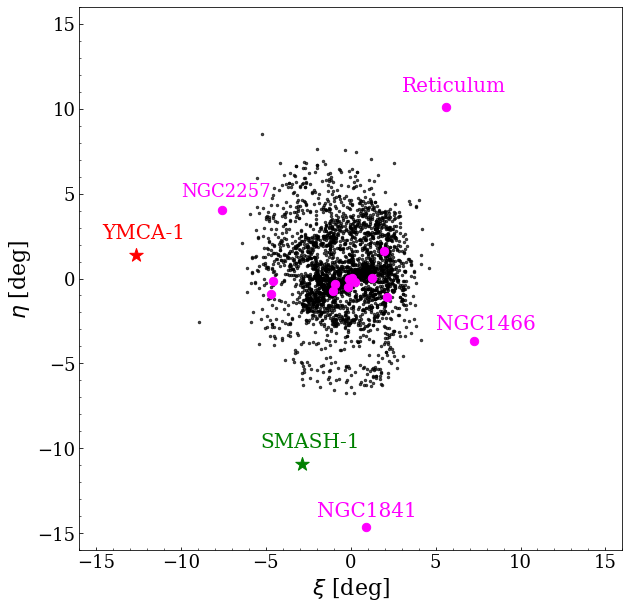}
    \caption{Relative position with respect to the LMC centre of the SCs present in the \citet{Bica-2008}'s catalog (black dots), old GCs (magenta points) and YMCA-1 and SMASH-1. Names of the outermost old GCs are also reported in the figure.}
    \label{fig:ymca1_relative_position}
\end{figure}

In Fig.~\ref{fig:simon_image} we display the luminosity ($M_V$\footnote{To estimate $M_V$ of YMCA-1 we first transformed $L_g$ and $L_i$ into $M_g$ and $M_i$. Then we used the following color transformation:  $V=g-(0.361\pm0.002)[(g-i)-1.0]-(0.423\pm0.001)$ with rms=0.024 mag. This equation was derived using several thousands of stars in the outskirts of the MCs having $V\,g\,i$ data from the APASS (The AAVSO Photometric All-Sky Survey) survey (https://www.aavso.org/apass).}) and half-light radius of YMCA-1 and SMASH-1, in comparison with those of old LMC GCs for which structural parameters were available in literature, and dwarf spheroidal galaxies (dSph) as well as UFDs from \citet{Simon2019}.  
The figure also shows the position of the old MW GCs whose parameters were taken from \citet{Baumgardt&Hilker2018} and \citet{Harris1996} (see also caption of Fig.~\ref{fig:simon_image}).
The proximity of YMCA-1 and SMASH-1 in this diagram is noticeable.  Both stellar systems lie in the region of the $M_V$~vs~$r_h$ space occupied by some peculiar faint MW GCs, such as AMR~4, Palomar~1, Koposov~1 and Koposov~2, and also near to objects with difficult classification but suspected to be at the faint end of the UFDs distribution. 
Even more interesting is the difference between YMCA-1 and the known old LMC GCs, which are located in a completely different locus of the $M_V$~vs~$r_h$ plane. They are several orders of magnitude more luminous and reside in the same parameter region occupied by the majority of the MW GCs.
Therefore, YMCA-1 and SMASH-1 might belong to a peculiar sub-class of stellar systems within the LMC whose properties are in between the classical GCs and the UFDs.
Unlike the more massive GCs, these low-dense objects are more sensitive to the external tidal fields and hence can be subject to complete disruption, which might explain the scarcity of these stellar systems in the LMC.
Indeed, \citet{Martin-2016a} concluded that SMASH-1 is experiencing an ongoing tidal disruption, based on its strong ellipticity and its estimated tidal radius.
Of course, other faint LMC-bound systems could still lay undiscovered in the outermost regions of the LMC.
Finally, Fig.~\ref{fig:ymca1_relative_position} shows the relative position of the LMC SCs with respect to the LMC centre. 
The picture reveals that YMCA-1 and SMASH-1 are among the farthest SCs ever detected around the LMC, but they are not spatially close, as the former is found to the East of the LMC, while the latter is in the South.
Moreover, YMCA-1 is superimposed (but not necessarily associated) to a substructure recently discovered in the north-est of the LMC \citep[i.e. the North-East Structure  or NES,][]{Gatto-2022}.\par
To summarize, YMCA-1 is likely an LMC old GC with features very similar to SMASH-1. 
Spectroscopic follow-up of both these interesting stellar systems can be very valuable to confirm their association with the LMC.
Until such spectroscopic confirmation is obtained, we cannot discard the less likely hypothesis that YMCA-1 (and possibly SMASH-1) is instead a MW remote GC. 

\section{Summary}

In this work we exploited the FORS2@VLT follow-up of YMCA-1, a new stellar system discovered within the context of the YMCA survey, placed at about 13\degr~from the LMC centre.
The deep catalogue obtained in this work ($g\sim26.5$~mag) allowed us to definitely confirm that YMCA-1 is a real physical stellar system.
The exploitation of its CMD by means of the automatic isochrone fitting package  {\tt ASteCA} \citep[][]{Perren-2015} and the analysis of its radial density profile, reveals that YMCA-1 is an old ($t = 11.7^{+1.7}_{-1.3}$~Gyrs), metal-intermediate ([Fe/H] $\simeq -1.12^{+0.21}_{-0.13}$~dex), low-mass (M $= 10^{2.45 \pm 0.02} M_{\odot}$) and compact (r$_{\rm h} \sim 3.5 \pm 0.5$ pc) stellar system. 
The new estimate of YMCA-1 distance modulus suggests that it could belong to the LMC  rather than to the MW halo as supposed on the basis of previous shallower VST data \citep[][]{Gatto-2021a}. Nonetheless, the uncertainties on its distance do not allow us to definitely rule out the possibility that YMCA-1 is indeed a satellite of the MW.
YMCA-1 properties are remarkably different from the ones of the 15 known old LMC GCs, as they are all very massive.
As far as we are aware, only SMASH-1 \citep[][]{Martin-2016a} exhibits properties similar to that of YMCA-1, an occurrence that might indicate they have a common origin.
Spectroscopic measurements of the brightest stars belonging to YMCA-1 (and SMASH-1) with the aim of obtaining their radial velocities and evaluating their metal abundances are pivotal to assess their LMC membership and to unveil the origin of these very interesting and rare stellar systems.

\begin{acknowledgments}

We warmly thank the anonymous referee for the helpful comments to this manuscript.
This work is based on Director Discretionary Time (DDT) kindly allocated by ESO to proposal 108.23LX.001.
M.G. and V.R. acknowledge support from the INAF fund "Funzionamento VST" (1.05.03.02.04).
This work has been partially supported by INAF through the “Main Stream SSH program" (1.05.01.86.28).
MRC acknowledges support from the European Research Council (ERC) under the European Union’s Horizon 2020 research and innovation programme (grant agreement no.682115).
This research was made possible through the use of the AAVSO Photometric All-Sky Survey (APASS), funded by the Robert Martin Ayers Sciences Fund and NSF AST-1412587.

\end{acknowledgments}





%

\vspace{10mm}
\facilities{Very Large Telescope (VLT), VLT Survey telescope \citep[VST,][]{Capaccioli&Schipani2011}}


\software{astropy \citep[][]{astropy-2013,astropy-2018}, astroquery \citep[][]{astroquery}, IRAF \citep[][]{IRAF-1-1986,IRAF-2-1993}, matplotlib \citep[][]{matplotlib}, scikit-learn \citep{scikit-learn}, scipy \citep[][]{Scipy}.}

\bibliography{unatesi}{}
\bibliographystyle{aasjournal}






\end{document}